\documentclass{elsart}
\usepackage{graphicx,amssymb}
\begin{document}
\begin{frontmatter}

\title{Outward-inward information flux in an opinion formation model 
on different topologies}
\author{A.O. Sousa\thanksref{label1}}
\thanks[label1]{{\it E-mail address:} sousa@ica1.uni-stuttgart.de, aosousa@gmail.com}
\address{Institute for Computer Physics (ICP), University of 
Stuttgart,\\ Pfaffenwaldring 27, 70569 Stuttgart, Germany.} 
\author{J. R. S\'{a}nchez\thanksref{label2}}
\thanks[label2]{{\it E-mail address:} jsanchez@fi.mdp.edu.ar}
\address{Fac. Ingenier\'{i}a - Univ. Nac. de Mar del Plata\\Justo 4302 - 
7600 Mar del Plata - Argentina.}
\begin{abstract} 
A simple model of opinion formation dynamics in which binary-state agents 
make up their opinions due to the influence of agents in a local neighborhood 
is studied using different network topologies. Each agent uses two different 
strategies, the Sznajd rule with a probability $q$ and the Galam majority 
rule (without inertia) otherwise; being $q$ a parameter of the system. 
Initially, the binary-state agents may have opinions (at random) against or 
in favor about a certain topic. The time evolution of the system is studied 
using different network topologies, starting from different initial opinion 
densities. A transition from consensus in one opinion to the other is 
found at the same percentage of initial distribution no matter which type 
of network is used or which opinion formation rule is used.
\end{abstract}

\end{frontmatter}

\section{Introduction}  
Recently, there has been a growing interest in the study of complex 
phenomena appearing in heterogeneous areas different from the traditional 
fields of research. In particular, one of those interesting fields turns 
to be the application of statistical physics methods to social phenomena. 
At present discussions of such heterogeneous aspects can be found in 
several reviews \cite{Weidlich2002,galam,stauffer1}. One of the 
sociological problems that attracts much attention is the construction of 
consensus out of some initial condition and different models have been 
introduced in order to simulate and analyze the dynamics of opinion formation 
processes \cite{galam,deffuant,stauffer1,sznajd,slanina,solomon,heider,stauffer2,vilone,cast,redner,stauffer3}. 

Mainly, opinion formation models are composed by interacting {\it agents}
usually represented by the state adopted by certain variables. Agents 
interact between each other forming a network with a certain topology. 
Models differ to each other in three aspects: the type of variables used to 
represent the opinion of the agent, the type of interaction imposed between 
the agents and the topology adopted. 

Here, in order to maintain certain degree of generality, the
dynamics of consensus formation is studied using different topologies. 
Each agent is represented by a spin state variable representing the opinion
in favor ($+1$) or against ($-1$) about a certain topic and the interaction
between the agents is randomly changed during the evolution between two
possible mechanisms, the Galam majority rule \cite{galam} and the 
Sznajd updating rule \cite{sznajd}. This last model is a successful 
Ising spin system describing a simple mechanism of making up decisions in a
closed community in which the opinion influence flows outward a given
group to the nearest individuals. In the Sznajd model a pair of nearest
neighbors convinces its neighbors to adopt the pair opinion if and only
if both members of the pair have the same opinion; otherwise the
pair and its neighbors do not change opinion. The Sznajd consensus model
has rapidly acquired importance in the new field of computational 
socio-physics \cite{Weidlich2002,stauffer1,sznajd,slanina,stauffer3,moreira,och,bonnekoh,chang,pseudo,sousa,schweitzer,sanchez}. 

Additionally, it is meaningful to mention that consensus models are set
up in complex networks with different topologies. The statistical 
properties of real-world social networks vary strongly, for example, 
the degree distribution can be single-scale, broad-scale or scale-free
\cite{amaral}. Due to the lack of a single model 
covering the topological features of social networks, we consider a few 
well established network models aiming to unveil the effect to different 
aspects of the topology:
\begin{itemize}
        \item standard one- and two- dimensional regular lattices;
        \item Watts-Strogatz small-world network \cite{watts};
        \item Erd\H os-R\'enyi random graph network \cite{random};
        \item Triad scale-free network (Barab\'asi-Albert-like 
model) \cite{barabasi,kertesz}.
\end{itemize}   

\section{The Model}     
On the system lattice, each site $i$ ($i = 1,2, \cdots,N$; where $N$ is 
the total number of sites-agents) carries a spin $s_i$, which has two possible 
directions, $s_i=+1$ or $s_i=-1$. The state of these spin-like variables can 
be considered to represent an agent or individual carryng one of two possible 
opinions, against or in favor about a certain topic: $s_i  =  + 1$ represents 
a positive opinion and $s_i=-1$, a negative one. Initially the opinions are 
distributed randomly, $s_i=+1$ with probability $p$ or $s_i = -1$, otherwise. 
\subsection{Opinion updating rules}
As stated above, during the time evolution of the model, each agent
can use  two different strategies in order to update its opinion state
variable $s_i^{ t}$,
\begin{itemize}
\item with probability $q$ the agent opinion is updated according to a kind 
of Sznajd rule \cite{stauffer1,sznajd,stauffer2,bonnekoh,pseudo,sousa}:
    \begin{itemize}
    \item {\bf square lattice and random graphs:} \underline {2 sites 
convincing rule}: For each site $i$ chosen, we select randomly one of 
its neighbors. If this selected neighbor has the same opinion as the site 
$i$, then all their neighbors follow the pair's opinion. Otherwise, 
nothing is done.
     \item {\bf usual 1d regular chain:} Two updating rules have been 
applied: a) the original Sznajd rule \cite{sznajd}: $s_{i-1}^{t}=s_{i+1}^{t}$ 
and $s_{i+2}^{t}=s_i^{t}$; b) 2 sites convincing rule.
      \end{itemize}
\item with probability $1-q$ the agent opinion is updated using 
a Galam-like majority rule without inertia in which individuals 
gather during their social life in {\sl meeting cells} of different 
sizes where they discuss about a topic until a final decision, in 
favor or against, is taken by the entire group.
\begin{equation}
  s_{i}^{t}= {\rm{sign}}\left(s_{i}^{t} + \frac{\sum\limits_{j \in 
\wedge_{i}} s_j^{t}}{k(i)}\right),
\end{equation}
where the ${\rm{sign}}(.)$ function takes the value $+1$
if the argument is positive and the value $-1$, $k(i)$ is the number of
vertices in the neighborhood $\wedge_i$ of site $i$: the nearest neighbors
in the usual one- and two- dimensional regular lattices and all nodes
connected to $i$ in the random graphs. In the regular lattices there is
never a tie. Although the random graphs have been constructed considering a
node connectivity to avoid such a tie, in the case it occurs, then the
individual randomly assumes an opinion $\pm 1$.
\end{itemize}
\subsection{Network topologies}
\noindent In this subsection we briefly describe the different topologies 
used, in addition to the standard one- and two- dimensional lattices. 
\subsubsection{Watts-Strogatz small-world network}
\noindent Social networks are far from being completely regular or 
completely random. For instance, it has been found that most real-life 
networks display some common characteristics the most important of which are 
called small-world effect and scale-free distribution. The recognition of 
small-world effect involves two factors: the clustering coefficient 
and the average shortest path length, a network is called a small-world 
network as long as it has small average shortest path length and large 
clustering coefficient \cite{watts,barrat}. One of the most 
well-known small-world models is Watts-Strogatz small-world network 
(WS model) \cite{watts}, which can be constructed by the following 
algorithm: the initial network is a one-dimensional lattice of $N$ 
sites, with periodic boundary conditions (i.e., a ring), each site being 
connected to $m$ nearest neighbors. We choose a vertex and the edge that 
connects it to its nearest neighbor in a clockwise sense. With probability 
$p_s$, we reconnect this edge to a vertex chosen uniformly at random 
over the entire ring, with duplicate edges forbidden; otherwise we leave 
the edge in place. This process repeats until one lap is completed 
and proceeding outward to more distant neighbors after each lap, until 
each edge in the original lattice has been considered once.
\subsubsection{Erd\H os-R\'enyi random graph}
The construction of the Erd\H os-R\'enyi random graph \cite{random} starts 
with a set of $N$ isolated vertices, then successive edges are randomly 
added with a probability $p_a$. In this way, the total number of edges is 
$m_{t}=p_{a}N(N-1)/2$ and the average number of neighbors of a node 
(degree or connectivity) is $\overline{m}=p_{a}(N-1)$. In the limit 
$N \rightarrow \infty$, the mean number of bonds per site can be 
approximated by $p_{a}N$ and a Poissonian connectivity distribution 
is observed. In our simulations, the graph has been built in such way 
that each node has at least $m$ links, although the coordination number 
of any particular vertex can be bigger than $m$ links ($m=4$ in our simulations).
\subsubsection{Barab\'asi-Albert and Triad scale-free network}
The most fashionable network presenting both properties, scale-free 
and small-world aspects, is the Barab\'asi-Albert scale-free network 
\cite{barabasi}. Although the BA network has successfully explained 
the scale-free nature of many networks, a striking discrepancy between 
it and real networks is that the value of the clustering coefficient 
varies fast with the network size $N$ and for large systems is typically 
several orders of magnitude lower than found empirically and it vanishes 
in the thermodynamic limit \cite{kertesz}. In social networks, for 
instance, the clustering coefficient distribution $C(k)$ exhibits a 
power-law behavior, $C(k) \propto k^{-\gamma}$, where $k$ is the node 
degree (number of neighbors) and $\gamma \approx 1$ (everyone in the 
network knows each other).
\begin{figure}[!h] 
\begin{center}
\includegraphics[width=6.95cm,height=5.85cm]{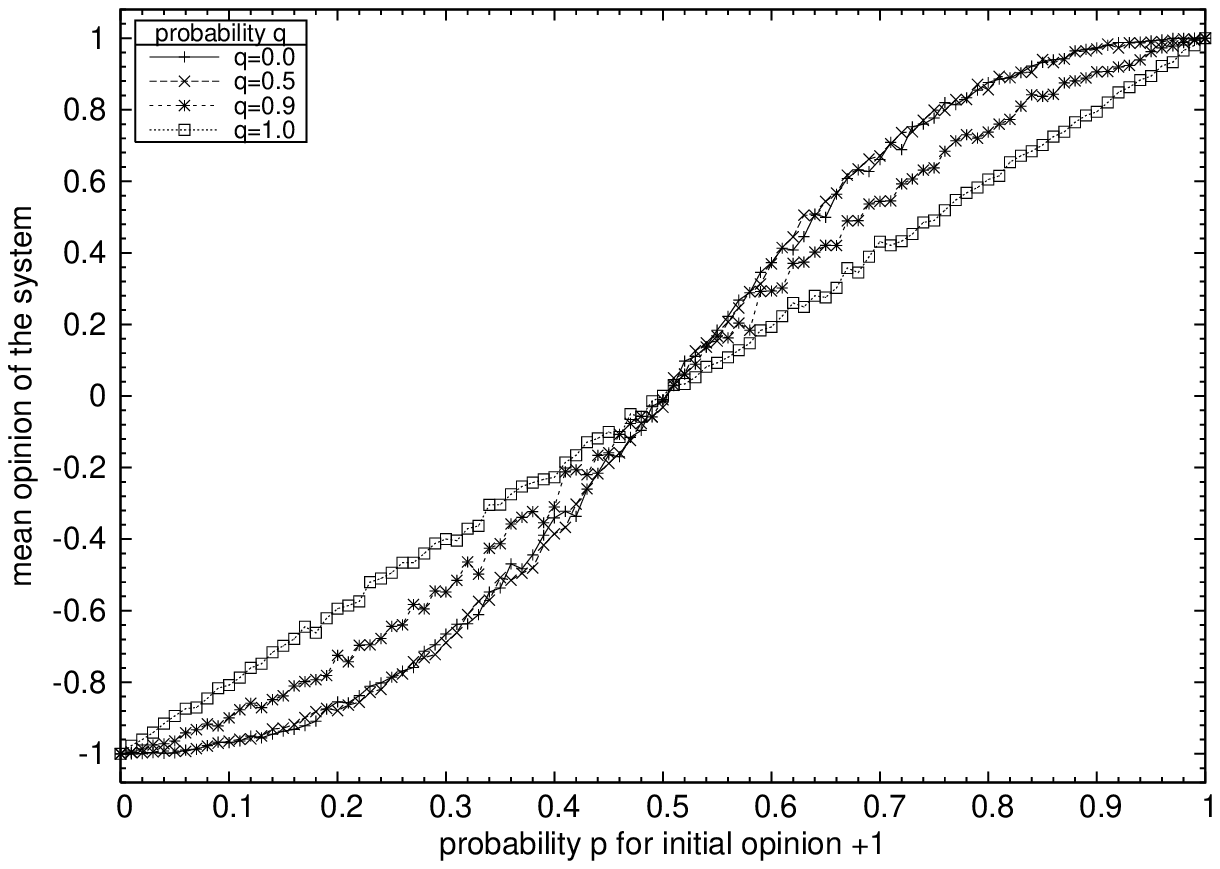}\hskip 0cm 
\includegraphics[width=6.95cm,height=5.85cm]{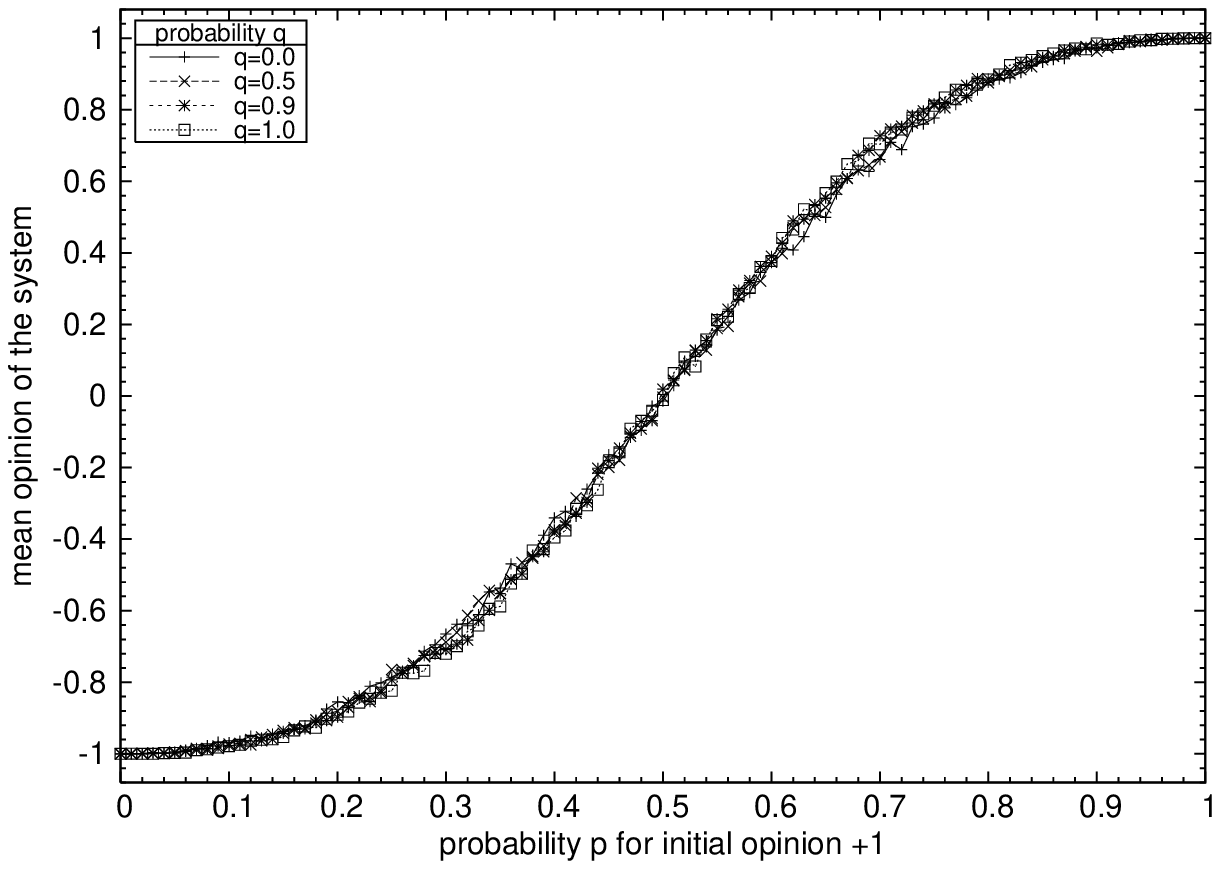}
\caption{The mean opinion density of the system as a function
  of the probability $p$ for initial opinion $+1$ for an one-dimensional 
lattice, different values of probability $q$, $N = 500$ and $n_{s} = 3000$. On
the left, original updating rule, and on the right, $2$ nodes convincing
rule.}
\label{fig:uni1}
\includegraphics[width=6.95cm,height=5.85cm]{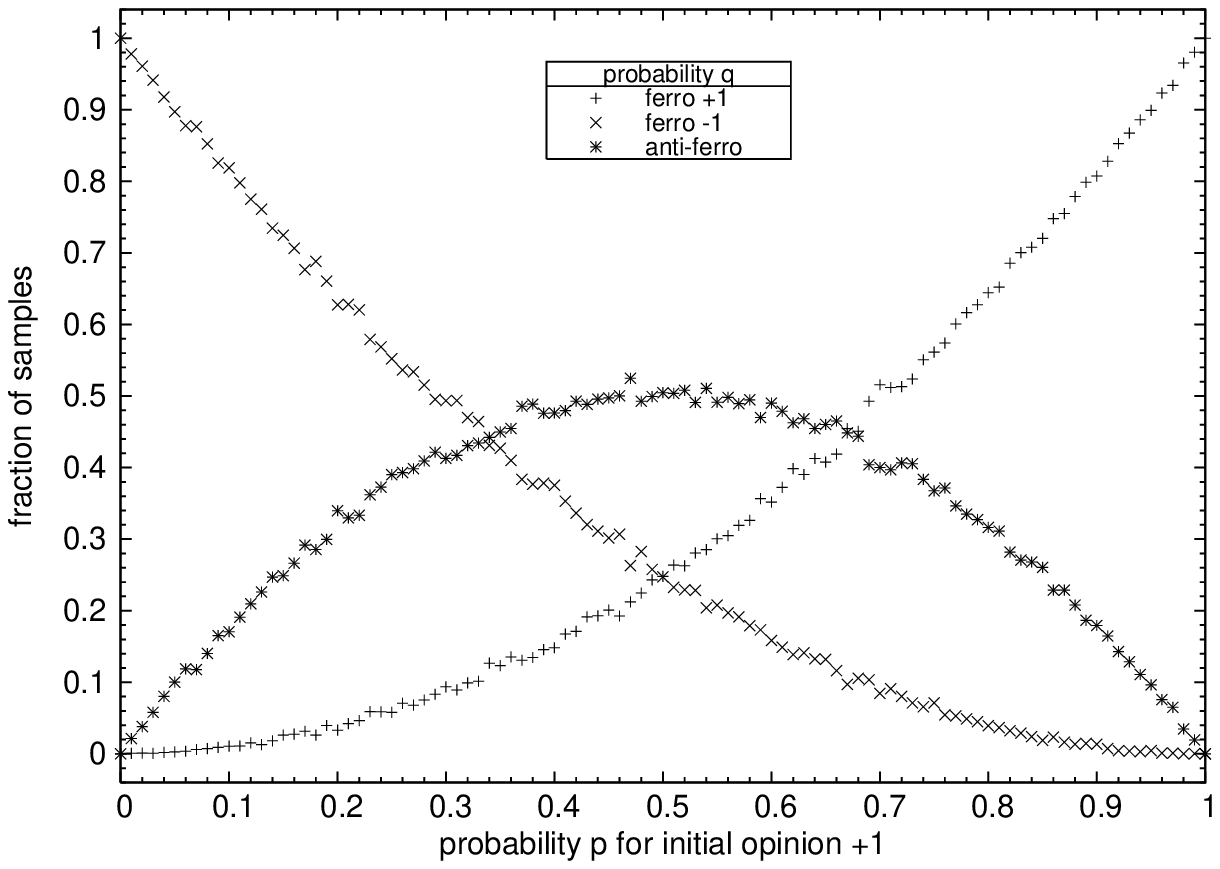}\hskip 0cm 
\includegraphics[width=6.95cm,height=5.85cm]{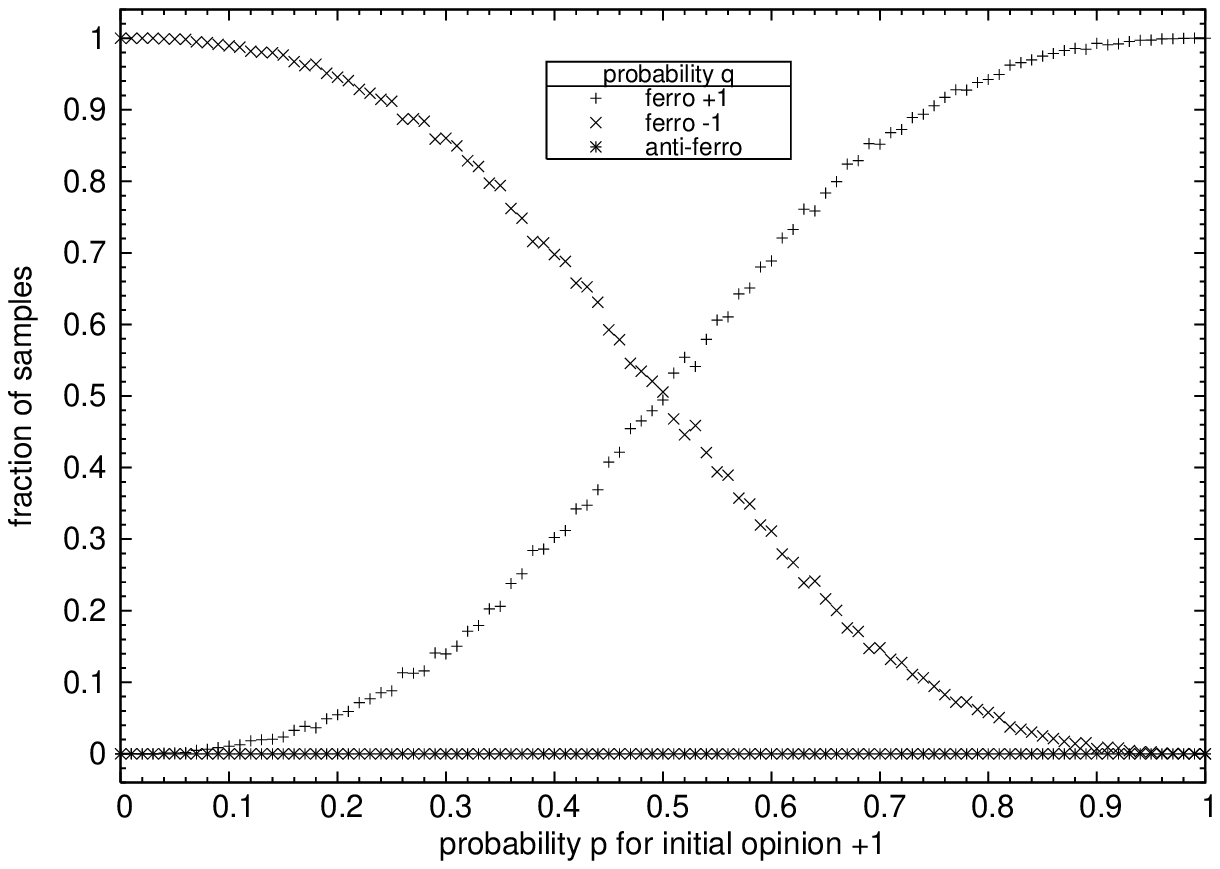}
\caption{Fraction of samples, out of 3000, which becomes ordered
  ferro-magnetically ( ``all up''(+) and ``all down''($\times$) ) and 
anti-ferromagnetically ($\ast$) as a function of the probability $p$ for 
initial opinion $+1$ for an one-dimensional lattice, $N = 500$ and 
$q=1.0$. On the left, original updating rule, and on the right, $2$ nodes 
convincing rule.}
\label{fig:uni2}
\end{center}
\end{figure}

This problem has been surmounted and scale-free models with high clustering 
coefficient have been investigated, by adding a triad formation step on the 
Barab\'asi-Albert prescription \cite{kertesz}.  The Barab\'asi-Albert 
network starts with a small number $m$ ($m=4$ in our simulations) of 
sites (agents, people) all connected with each other. Then a large number 
$N-m$ of additional sites is added as follows: first, each new node 
(node $i$) performs a preferential attachment step, i.e, it is attached 
randomly to one of the existing nodes (node $j$) with a probability 
proportional to its degree; then follows a triad formation step with a 
probability $p_t$: the new node $i$ selects at random a node in the 
neighborhood of the one linked to in the previous preferential attachment 
step (node $j$). If all neighbors of $j$ are already connected to $i$, 
then a preferential attachment step is performed (``friends of friends 
get friends''). In this model, the original Barab\'asi-Albert network 
corresponds to the case of $p_{t}=0$. It is expected that a nonzero 
$p_t$ gives a finite nonzero clustering coefficient as $N$ is 
increased, while the clustering coefficient goes to zero when 
$p_{t}=0$ (the BA scale-free network model) \cite{kertesz}. Indeed, 
the clustering coefficient increases as the probability $p_t$ and 
$m$ increase.
\section{Results}
At every time step $t>0$, all the individuals $N$ are randomly visited 
and updated (a random list of nodes assures that each node is 
reached exactly once) by following the stated rules in Section 2.1. 
If a full consensus (all individuals have the same opinion) or if 
the maximum number $t_{max}=100000$ of iterations is reached, the 
simulation ends. The curves presented here correspond to the results 
averaged over $n_s$ samples. 

In Fig.\ref{fig:uni1} we present the mean opinion of the system as 
a function of the probability $p$ for the one dimensional model when 
different values of probability $q$ are considered. On the left side, 
the results are from using majority \cite{galam} and the original 
Sznajd \cite{sznajd} rules. As we can see, the mean 
opinion (or final magnetization $\rm m_f$) is given by $\rm m_f=2p-1$ 
when only the Sznajd rule ($q=1.0$) is applied and as $q$ increases 
$\rm m_f$ quickly approaches a static value 
$\rm m_f=m_{o}(3/(2+m_{o}^{2}))^{3/2}$, that depends only on the 
initial magnetization $m_o=2p-1$. On the right side, it can be 
noticed that the system exhibits the same behavior for any value 
of probability $q$. Moreover, from this plot, we find that in both 
updating rules, the 2 nodes convincing and the majority rule, the system 
always evolves to a ferromagnetic consensus (either all up or all down) 
and that they provide the same probability distribution for the system 
to reach a fixed point with all spins up or all spins down 
(Fig.\ref{fig:uni2}, right), except in the anomalous case of an
anti-ferromagnetic initial state ($+-+-+-$). Nevertheless, when the 
system updates according to the original rule, besides the two 
ferromagnetic states, an anti-ferromagnetic fixed point is also reached 
(Fig.\ref{fig:uni2}, left). These results are in good agreement with 
previous studies on Sznajd, majority and minority models 
\cite{galam,sznajd,redner,schweitzer}. Very recently, 
a modification of the Sznajd model in order to avoid the unrealistic 
anti-ferromagnetic fixed point was proposed in Ref.\cite{sanchez}.

\begin{figure}[!h] 
\begin{center}
\includegraphics[width=6.95cm,height=5.85cm]{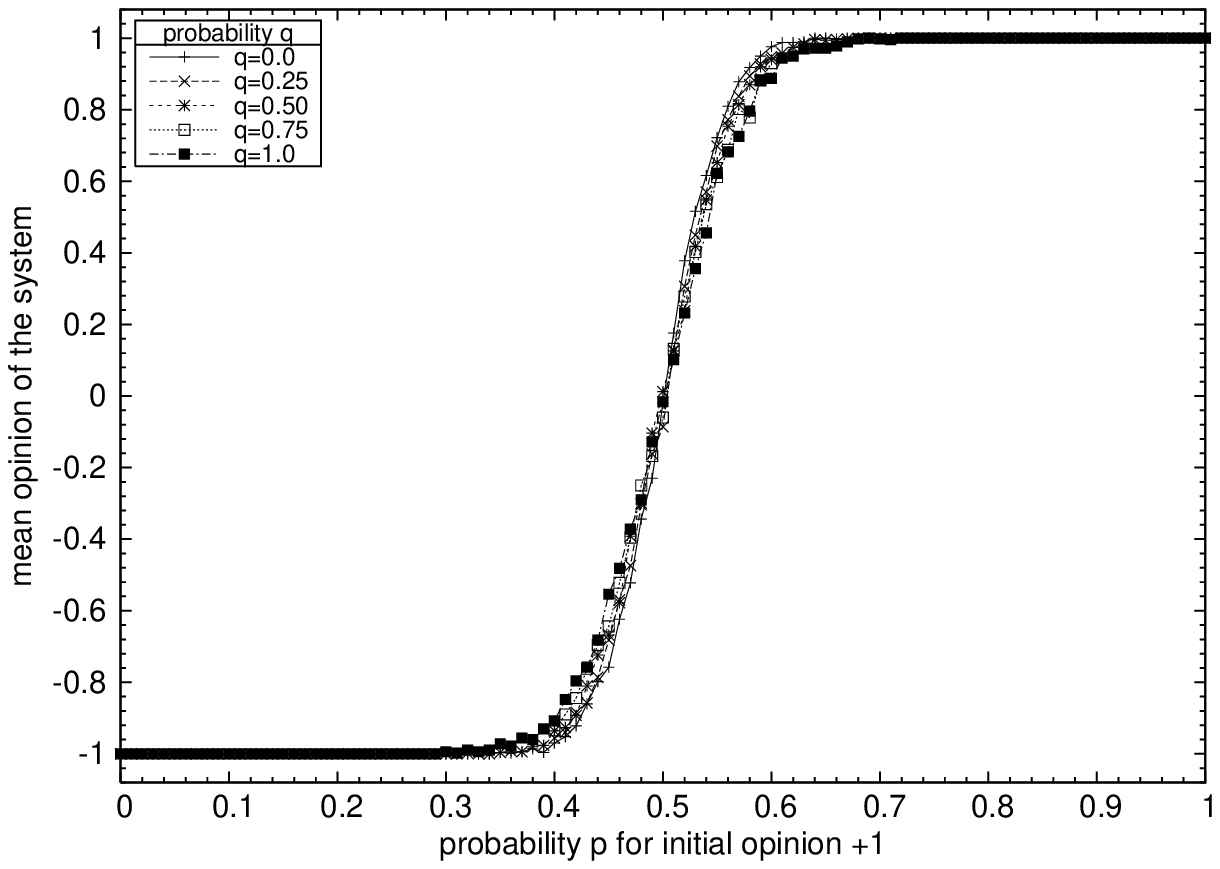}\hskip 0cm 
\includegraphics[width=6.95cm,height=5.85cm]{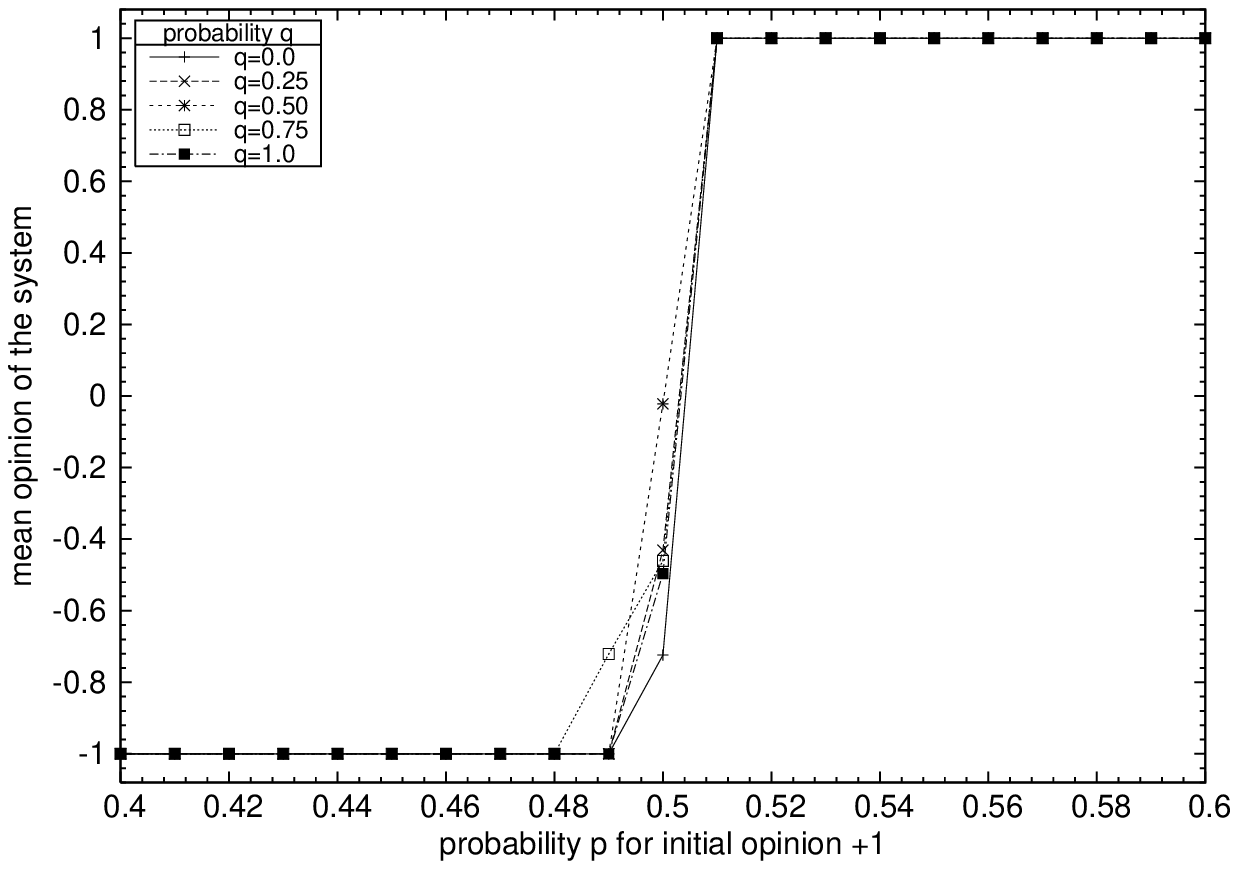}
\caption{As Fig. \ref{fig:uni1}, but for the square lattice. On the 
left, $N=400$ and $n_{\rm s}=1000$, and on the right side, $N=490000$ 
and $n_{\rm s}=1$.}
\label{fig:sq}
\end{center}
\end{figure}

In Fig. \ref{fig:sq}, we observe the mean opinion density of the system as a 
function of the probability $p$ for initial opinion $+1$ for a square 
lattice, as well as for different values of probability $q$. From this 
figure, one can observe a phase transition: concentration $p>1/2$ lead 
to full consensus $+1$ and concentrations $p<1/2$ to full consensus 
$-1$ for large enough systems. Since in a finite network, of course, 
phase transitions are never sharp, and thus the transition is indicated 
numerically by a slope (Fig.\ref{fig:sq} left)
becoming the steeper the larger the network size is (Fig.\ref{fig:sq}
right). In an infinite network, one would get a sharp step function for the
mean opinion of the systems versus the initial concentration $p$ of opinion
$+1$. Moreover, the results are the same independently of the value of the
probability $q$, i.e., to update the system following Sznajd recipe or
majority rule provides the same behavior and consensus threshold $p=1/2$.
\begin{figure}[!h] 
\begin{center}
\includegraphics[width=6.95cm,height=5.85cm]{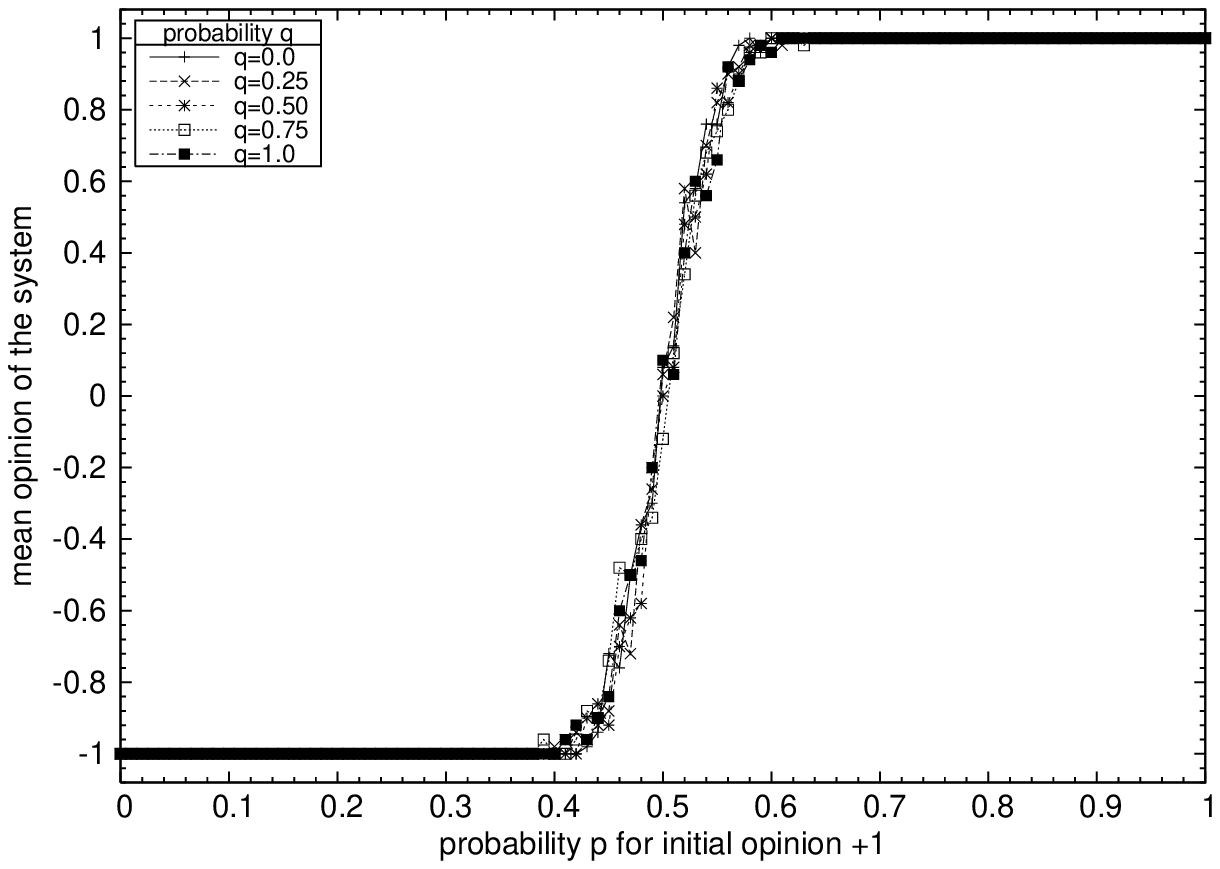}\hskip 0cm 
\includegraphics[width=6.95cm,height=5.85cm]{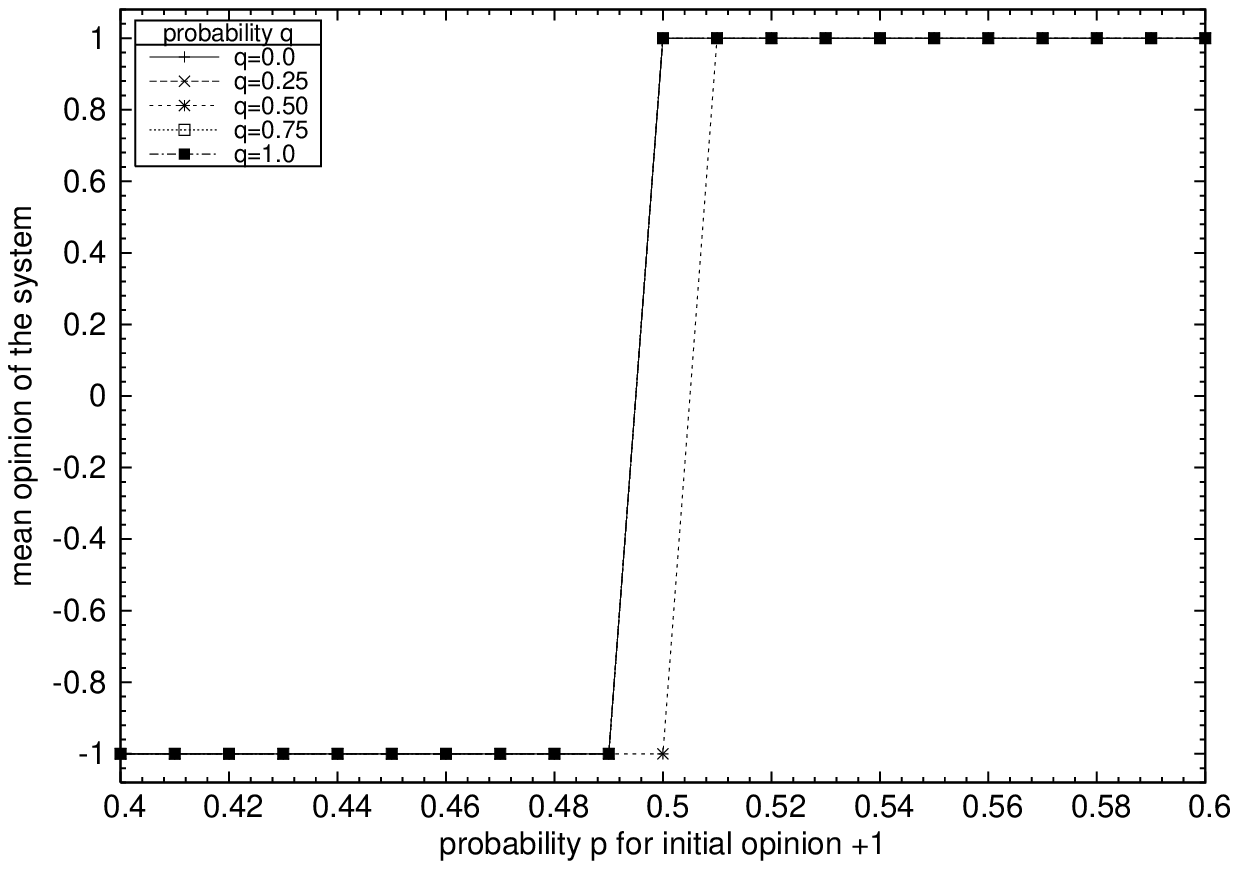}
\caption{As Fig. \ref{fig:uni1}, but for Erd\H os-R\'enyi random
  graph. On the left, $N=500$ and $n_{\rm s}=100$, and on
  the right side, $N=500000$ and $n_{\rm s}=1$.}
\label{fig:erdos}
\end{center}
\end{figure}

Fig. \ref{fig:erdos} shows the mean opinion density of the system as a 
function of the probability $p$ for initial opinion $+1$ for the 
Erd\H os-R\'enyi random graph. As we can see, it presents the same pattern 
as in Fig. \ref{fig:sq}, so that the consensus threshold is $1/2$ and 
there is no difference related to the chosen dynamics for updating the 
system: Sznajd or majority rule.  
\begin{figure}[!h] 
\begin{center}
\includegraphics[width=6.95cm,height=5.85cm]{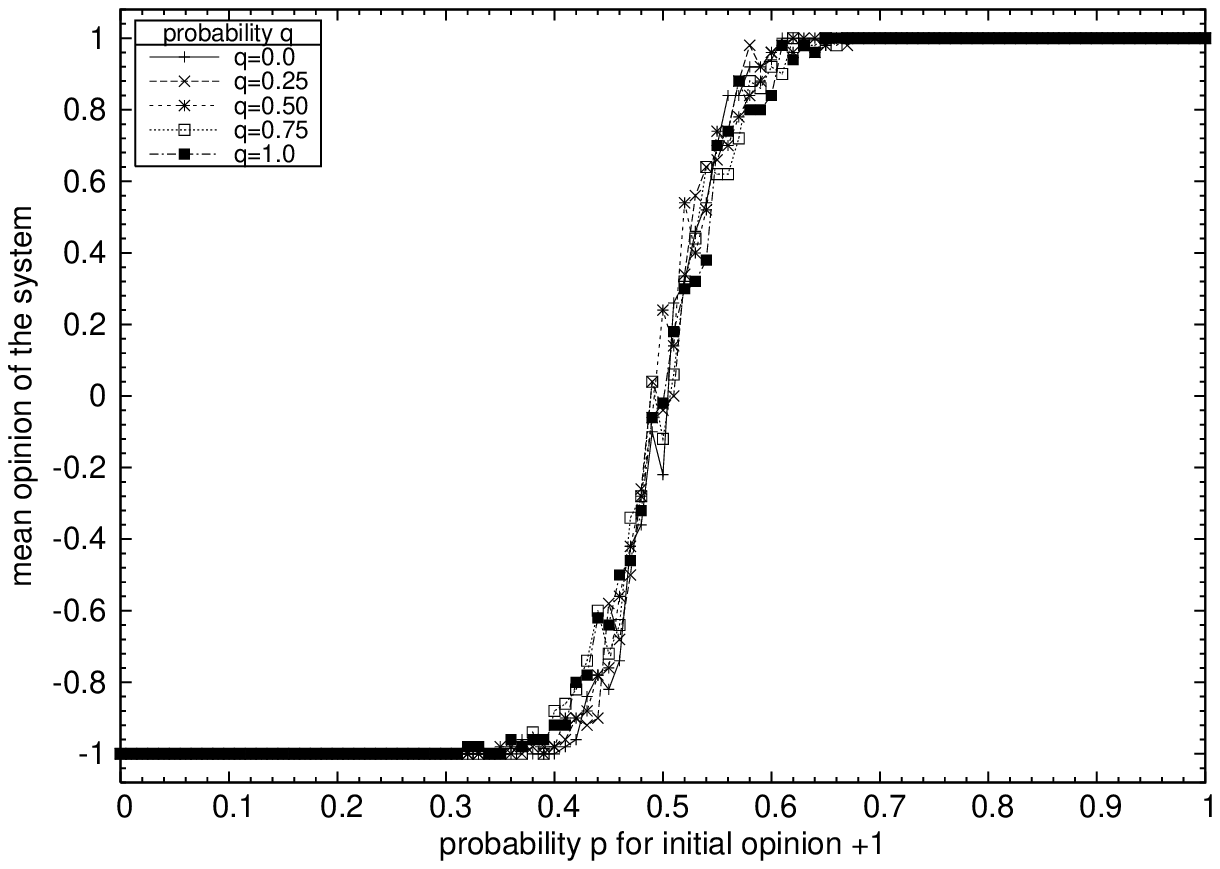}\hskip 0cm 
\includegraphics[width=6.95cm,height=5.85cm]{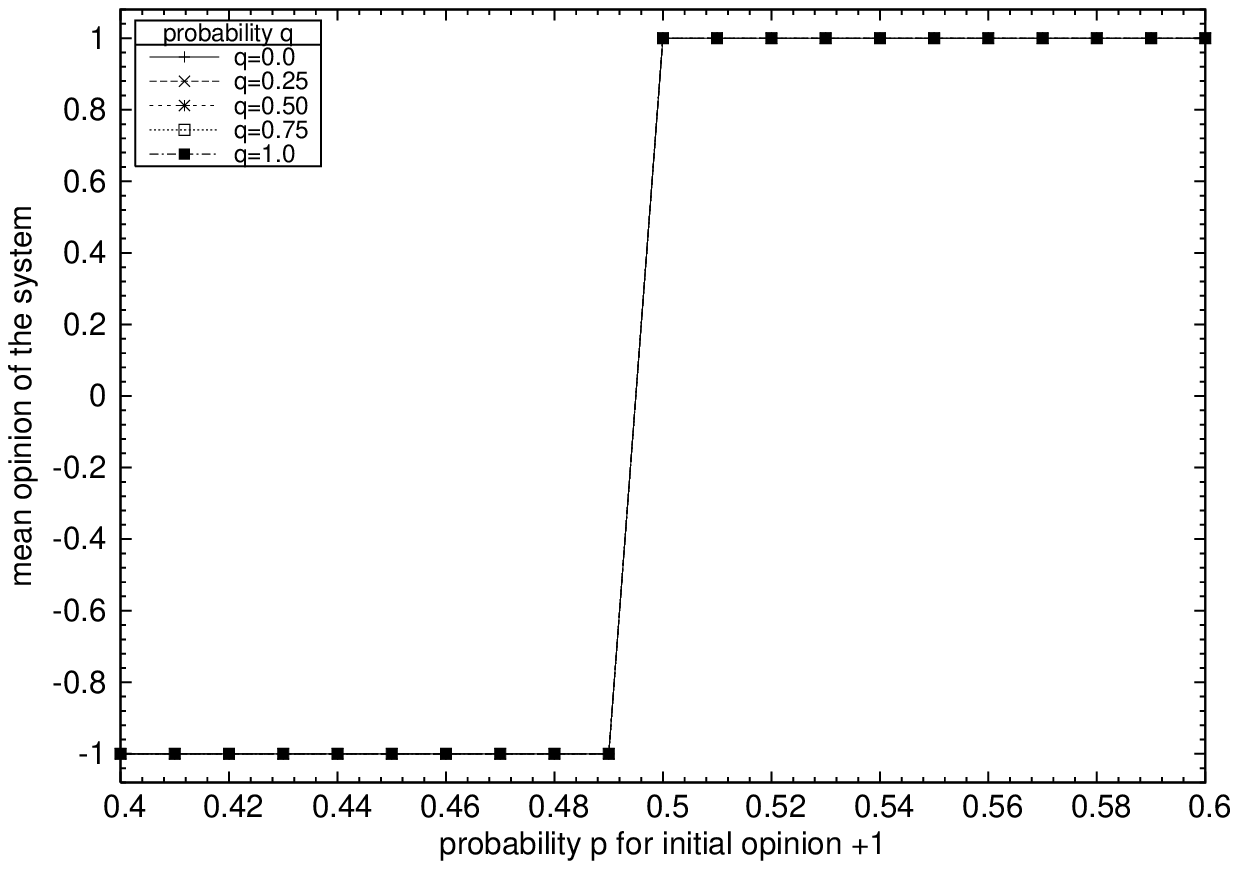}
\caption{As Fig. \ref{fig:uni1}, but for Watts-Strogatz
  small-world network. On the left, $N=500$ and $n_{\rm s}=100$, and on
  the right side, $N=500000$ and $n_{\rm s}=1$. Others parameters: 
$p_{s}=0.1$ and $m=2$}
\label{fig:watts}
\end{center}
\end{figure}

In Fig. \ref{fig:watts}, we show the same as in Fig.\ref{fig:sq}, but 
for the Watts-Strogatz small-world network. From this figure, it can be 
observed that the consensus threshold is also $1/2$ and the chosen 
updating dynamics does not exert any influence on the results, as already 
noticed for the square lattice and the Erd\H os-R\'enyi random graph.
\begin{figure}[!h] 
\begin{center}
\includegraphics[width=6.95cm,height=5.85cm]{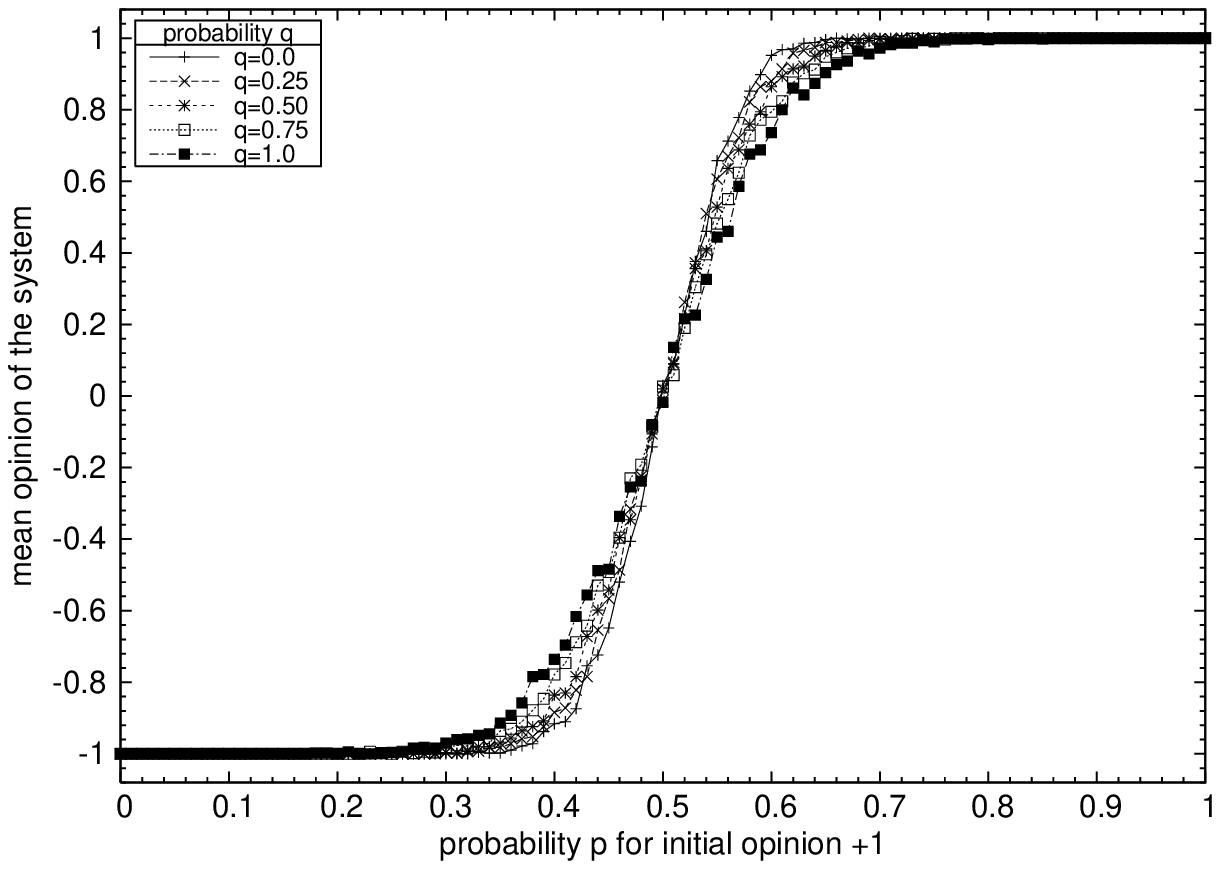}\hskip 0cm 
\includegraphics[width=6.95cm,height=5.85cm]{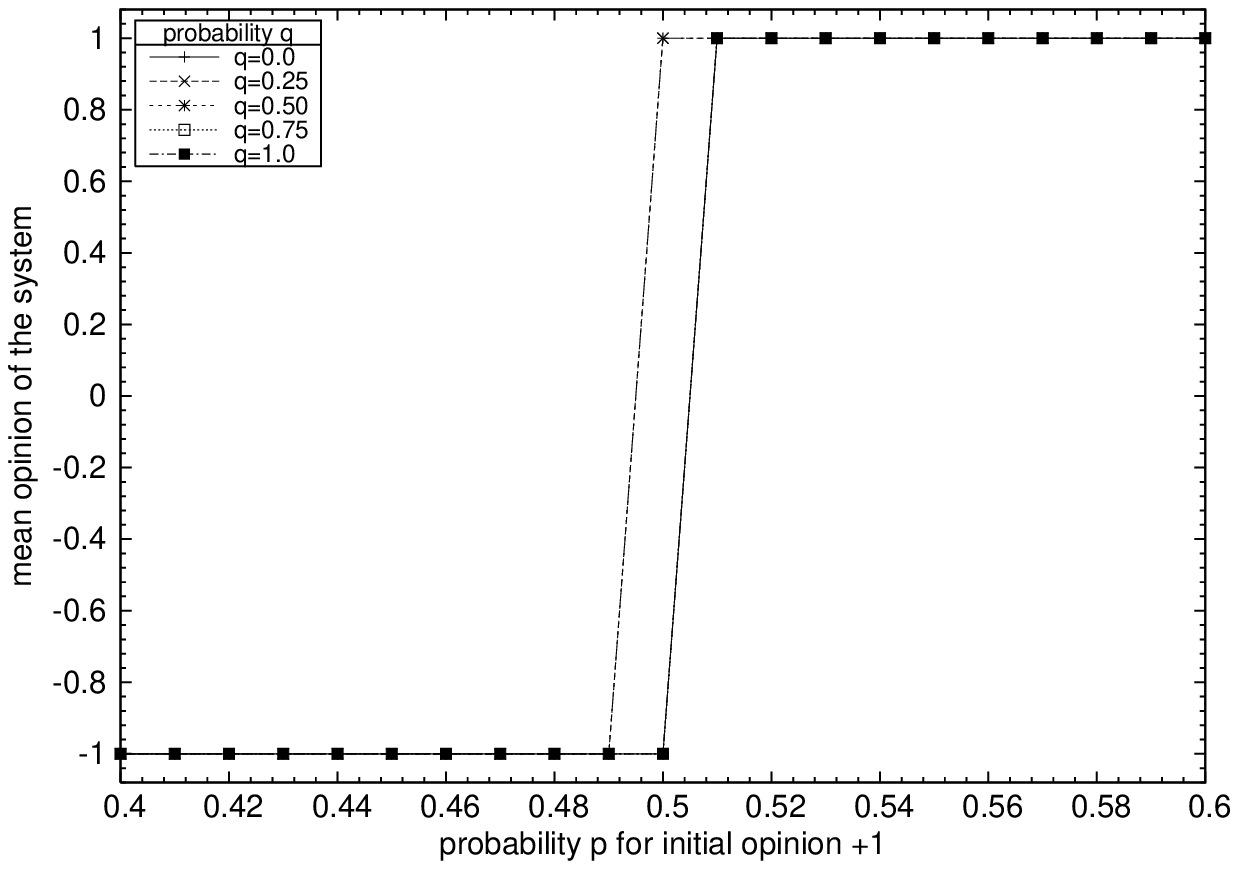}
\caption{As Fig. \ref{fig:uni1}, but for the Triad scale-free
  network. On the left, $N=500$ and $n_{\rm s}=1000$, and on
  the right side, $N=500000$ and $n_{\rm s}=1$. $p_{a}=m/N$, with $m=4$  }
\label{fig:triad}
\end{center}
\end{figure}

Fig. \ref{fig:triad} presents the mean opinion density of the system 
as a function of the probability $p$ for initial opinion $+1$ for the 
Triad scale-free network. The same consensus threshold $p=1/2$ and 
results completely independent of the adopted updating dynamics are 
once more observed. 

The phase transition at $p_{c}=1/2$ here observed for the square lattice and
for the random graphs does not exist in one dimension \cite{sznajd} or 
when a single site \cite{och} (instead of a pair or plaquette) on the square 
lattice \cite{stauffer2} convinces its neighbors, although it has been also 
found on the square lattice when a plaquette or a neighboring pair 
persuades its neighbors \cite{stauffer2}, on a correlated-diluted square 
lattice \cite{moreira}, on a triangular and simple cubic lattice if a pair 
convinces its $8$ (or $10$, respectively) neighbors \cite{chang}, on the 
Barab\'{a}si-Albert network \cite{bonnekoh}, on a triad network \cite{sousa} 
and on a pseudo-fractal network \cite{pseudo}.

\section{Conclusion}
In this article, we have investigated a simple model of opinion formation 
in which binary-state agents evolve due to the influence of agents 
in a local neighborhood by two different updating rules: with 
probability $q$, the agents adopt the Sznajd rule and the majority 
rule with probability $1-p$. In one dimension, for $q>0$ the system 
always evolves to three fixed points: two of which refer to ferromagnetism 
(``all up'' or ``all down'') and one to anti-ferromagnetism. As $q$ 
increases, the magnetization quickly approaches a fixed values that 
depends only on the initial magnetization. On the other hand, we have also 
obtained that the 2 nodes convincing rules corresponds to a kind of majority 
rule in which only a ferromagnetic consensus is observed with similar
probability distribution for a consensus +1 or consensus -1. Moreover, 
taking into account more realistic topologies, the interplay between 
majority and Sznajd rule leads the system to a phase transition as a
function of the initial concentration of $S=+1$ opinions at $p_{c}=1/2$:
For $p < 1/2$ all samples end up with $S=-1$, and for $p > 1/2$  they all
end up in the other fixed point $S=+1$, for large enough systems. Indeed, 
such consensus threshold $p_c$ of the opinion dynamics is not specific of
the particular graph one uses to describe the social relationships between
individuals, since it takes only one possible value $p_c=1/2$. Moreover, 
no matter whether the direction of the information flux is 
inward ($q=0$, majority rule) or outward ($q=1$, Sznajd rule) the 
system provides the same 
behavior. In fact, our results represent a natural outgrowth of recent 
works \cite{galam} on the interplay between the majority rule and Sznajd 
model which claim that the Sznajd model represents a particular case 
of a class of Galam majority model, that shows the same threshold 
value $p_c$.
\section{Acknowledgements}
We thank D. Stauffer, S. Galam and K. Malarz for many helpful discussions, 
advice and a criticial reading of the manuscript.

\end{document}